\documentclass[aps,prl,twocolumn,showpacs,superscriptaddress]{revtex4-1}
\usepackage{enumerate}
\usepackage{graphicx}
\usepackage{amsmath}
\usepackage{amsfonts}
\usepackage{amssymb}
\newcommand{\ket}[1]{\left| #1 \right\rangle}
\newcommand{\bra}[1]{\left\langle #1 \right|}

\begin{document}
\today
\title{Comment on `Detecting non-Abelian geometric phases with three-level $\Lambda$
systems'}
\author{Marie Ericsson}
\affiliation{Department of Quantum Chemistry, Uppsala University,
Box 518, Se-751 20 Uppsala, Sweden}
\author{Erik Sj\"oqvist}
\affiliation{Department of Quantum Chemistry, Uppsala University,
Box 518, Se-751 20 Uppsala, Sweden}
\affiliation{Centre for Quantum Technologies, National University of Singapore,
3 Science Drive 2, 117543 Singapore, Singapore}
\date{\today}
\begin{abstract}
In their recent paper, Yan-Xiong Du {\it et al.} [Phys. Rev. A {\bf 84}, 034103 (2011)] claim to 
have found a non-Abelian adiabatic geometric phase associated with the energy eigenstates of 
a large-detuned three-level $\Lambda$ system. They further propose a test to detect the 
non-commutative feature of this geometric phase. On the contrary, we show that the non-Abelian 
geometric phase picked up by the energy eigenstates of a $\Lambda$ system is trivial in the 
adiabatic approximation, while, in the exact treatment of the time evolution, this phase is 
very small and cannot be separated from the non-Abelian dynamical phase acquired along 
the path in parameter space.
\end{abstract}
\maketitle
In a recent paper, Yan-Xiong Du {\it et al.} \cite{du11} claim to have found a measurable
non-Abelian geometric phase (GP) in adiabatic evolution of a three-level $\Lambda$ system. 
At first glance, this 
is a somewhat paradoxical claim since a nontrivial non-Abelian GP in adiabatic 
evolution requires degenerate energy eigenstates \cite{wilczek84}, such as the two 
dark states of a tripod system \cite{unanyan99}, while the energy eigenstates of a 
$\Lambda$ system are all non-degenerate. Therefore, the idea of Ref. \cite{du11} 
is to look at the $\Lambda$ system in the large detuning regime, where two of the 
energy eigenstates become nearly degenerate and may therefore pick up a non-Abelian 
adiabatic GP. 

Contrary to the claim in Ref. \cite{du11}, we show here that the 
non-Abelian GP of a large-detuned $\Lambda$ system in the adiabatic approximation 
is trivial. We further show that the non-Abelian GP picked up by this pair of nearly 
degenerate energy eigenstates in the exact treatment of the time evolution is small 
and cannot be separated from the non-Abelian dynamical phase acquired along the path 
in parameter space. This latter result implies that the non-commutative feature of this 
GP cannot be detected in an experiment. 

Following Ref. \cite{du11}, the considered system is a cold atomic gas, with each atom having an 
internal three-level $\Lambda$-type configuration. The two ground state levels $\ket{1}$ 
and $\ket{2}$ are coupled to an excited state $\ket{3}$ by two laser beams associated with 
Rabi frequencies $\Omega_1 = \Omega \sin \theta e^{i\varphi}$ and $\Omega_2 = \Omega 
\cos \theta$. Here, $\theta,\varphi$, and $\Omega$ are slowly varying parameters in time. 
The Hamiltonian in a frame that rotates with the laser fields reads
\begin{eqnarray}
H & = & -\hbar \left( \Omega \sin \theta e^{i\varphi} \ket{1} \bra{3} +
\Omega \cos \theta \ket{2} \bra{3} \right.
\nonumber \\
 & & \left. + 2\Delta \ket{3} \bra{3} \right) + \textrm{H.c.} ,
\label{eq:hamiltonian}
\end{eqnarray}
where rapidly oscillating counter-rotating terms have been neglected (rotating wave 
approximation). The two laser fields are assumed to have the same detuning 
$\Delta$. Diagonalizing $H$ yields the exact eigenvectors
\begin{eqnarray}
\ket{\psi_1} & = & \cos \theta \ket{1} - \sin \theta e^{-i\varphi} \ket{2} ,
\nonumber \\
\ket{\psi_2} & = & \cos \gamma \left(\sin \theta e^{i\varphi} \ket{1} +
\cos \theta \ket{2} \right) - \sin \gamma \ket{3} ,
\nonumber \\
\ket{\psi_3} & = & \sin \gamma \left(\sin \theta e^{i\varphi} \ket{1} +
\cos \theta \ket{2} \right) + \cos \gamma \ket{3} ,
\label{eq:exactvectors}
\end{eqnarray}
where $\tan \gamma = \left( \sqrt{\Delta^2 + \Omega^2} - \Delta \right) / \Omega$.
The corresponding exact energy eigenvalues are $\lambda_1 = 0$, $\lambda_2 = -\hbar
\left( \Delta - \sqrt{\Delta^2 + \Omega^2} \right)$, and $\lambda_3 = -\hbar \left( \Delta +
\sqrt{\Delta^2 + \Omega^2} \right)$.

In the limit of large positive detuning $\Delta \gg |\Omega|$, the first two eigenstates become 
nearly degenerate. Furthermore, $\sin\gamma \approx 0$ in this limit, which implies 
that the second and third eigenvectors can be approximated as $\ket{\psi_2} \approx 
\sin \theta e^{i\varphi} \ket{1} + \cos \theta \ket{2}$ and $\ket{\psi_3} \approx \ket{3}$. 
Thus, $\ket{\psi_3}$ is decoupled from the nearly degenerate energy eigenstates 
$\ket{\psi_1},\ket{\psi_2}$. These two latter states span a two-dimensional subspace 
$\mathcal{S}$ of the three-level system. However, $\mathcal{S}$ is independent of the 
adiabatic parameters $\theta,\varphi$ \cite{remark} and the corresponding matrix-valued 
Wilczek-Zee vector potential $A_{ab\mu} =\bra{\psi_a} \partial_{\mu} \ket{\psi_b}$ \cite{wilczek84} 
must therefore be a pure gauge. In other words, the non-Abelian gauge field $F_{\mu\nu} = 
\partial_{\nu} A_{\mu} - \partial_{\mu} A_{\nu} - [A_{\mu},A_{\nu}]$ must vanish.
This can be verified by explicit calculation. We first find 
\begin{eqnarray}
A_{\theta} & \approx & i\sigma_y \cos \varphi + i \sigma_x \sin \varphi ,
\nonumber \\
A_{\varphi} & \approx & -i \sigma_z \sin^2 \theta + i \sigma_x \sin \theta \cos \theta \cos \varphi
\nonumber \\
 & & - i \sigma_y \sin \theta \cos \theta \sin \varphi ,
\label{eq:Aapprox}
\end{eqnarray}
where $\sigma_x,\sigma_y,\sigma_z$ are standard Pauli matrices defined in the  
$\ket{1},\ket{2}$ basis. Notice the sign in the first term on the right-hand side of the expression
for $A_{\theta}$, which is opposite to that of the corresponding expression in Eq. (7) of Ref.
\cite{du11}. By using Eq. (\ref{eq:Aapprox}), one verifies that $F_{\theta\varphi}\approx 0$. 
On the other hand, by using Eq. (7) of Ref. \cite{du11}, one obtains a non-vanishing 
$F_{\theta\varphi}$, which contradicts the fact that $\mathcal{S}$ is independent of the 
adiabatic parameters $\theta,\varphi$. The origin of these conflicting results regarding the 
gauge field $F_{\theta\varphi}$ is exactly the sign difference in the $\sigma_y$ term of 
$A_{\theta}$ noted above. 

Since the gauge field vanishes, it follows that the non-Abelian GP is trivial, contrary to the 
claim in Ref. \cite{du11}. To verify this explicitly, we need to choose a single-valued basis 
$\ket{\eta_a}$ of the two-dimensional subspace $\mathcal{S}$, in terms of which the vector 
potential $\mathcal{A}_{ab\mu} = \bra{\eta_a} \partial_{\mu} \ket{\eta_b}$ and the non-Abelian 
GP $U_g = {\bf P} e^{-\oint \mathcal{A}_{\mu} dx^{\mu}}$ can be calculated. Two 
different single-valued bases $\ket{\eta_a}$ and $\ket{\eta_a'}$ can be related by a 
single-valued unitary $2\times 2$ matrix $V$ as $\ket{\eta_a'} = \sum_b \ket{\eta_b} 
V_{ba}$. The change $\ket{\eta_a} \mapsto \ket{\eta_a'}$ induced by $V$ is a gauge 
transformation. The corresponding non-Abelian GPs $U_g$ and $U_g'$ are related as  
$U_g' = V_0 U_g V_0^{\dagger}$, where $V_0$ generates the transformation between 
the initial bases \cite{kult06}. Now, by choosing $\ket{\eta_a} = \ket{a}$, we obtain 
$\mathcal{A}_{ab\mu} = \bra{a} \partial_{\mu} \ket{b} = 0$, which implies $U_g = \hat{1}$ 
and $U_g' = V_0 \hat{1} V_0^{\dagger} = \hat{1}$, where $\hat{1}$ is the $2\times 2$ 
identity matrix and we have used that $V_0$ is unitary. This holds 
for any single-valued basis; in particular, if we choose $\ket{\eta_a'} = \ket{\psi_a}$, then 
$V = \hat{1} \cos \theta + i\sigma_y \sin \theta \cos \varphi + i\sigma_x \sin \theta 
\sin \varphi$ and $\mathcal{A}_{\mu}$ becomes identical to $A_{\mu}$ in Eq. (\ref{eq:Aapprox}). 
This demonstrates that any closed path integral of a matrix-valued gauge potential for 
this system, such as $A_{\mu}$ in Eq. (\ref{eq:Aapprox}), gives rise to a trivial non-Abelian 
geometric phase. 

One may wonder whether the vanishing gauge field is an artifact of the neglect of 
$- \sin \gamma \ket{3}$ in $\ket{\psi_2}$ in the large detuning limit. Indeed, by including 
this term in the calculation, the gauge field $F_{\theta\varphi}$ turns out to be non-vanishing, 
and this may potentially cause a measurable non-Abelian GP effect. However, as we show next, 
this gauge field is very small and the effect of the non-Abelian GP can therefore be neglected 
in the large-detuning limit. Furthermore, it turns out that the dynamical $2\times 2$ matrix 
$D_{kl}=\bra{\psi_k} H \ket{\psi_l}$, $k,l=1,2$, does not commute with the Wilczek-Zee vector 
potential, which implies that the non-Abelian dynamical phase ${\bf T} e^{-(i/\hbar) 
\int_0^{\tau} D dt}$ cannot be canceled. In other words, the non-commutative feature of 
the non-Abelian GP of the energy eigenstates in the $\Lambda$ system cannot be detected 
in an experiment. 

First, we prove that the effect of the non-Abelian GP is negligible in the large detuning limit. 
By using the exact expressions for $\ket{\psi_1},\ket{\psi_2}$ in Eq. (\ref{eq:exactvectors}),
we obtain the Wilczek-Zee potential
\begin{eqnarray}
A_{\theta}  & = &  (i\sigma_y \cos \varphi + i \sigma_x \sin \varphi) \cos \gamma,
\nonumber \\
A_{\varphi}  & = & -i \sigma_z \sin^2 \theta - i \frac{1}{2} (\hat{1}-\sigma_z) \sin^2 \theta 
\sin^2 \gamma
\nonumber \\
 & & + i \sigma_x \sin \theta \cos \theta \cos \varphi \cos \gamma
\nonumber \\
 & & - i \sigma_y \sin \theta \cos \theta \sin \varphi \cos \gamma,
\label{eq:Aexact}
\end{eqnarray}
and corresponding gauge field 
\begin{eqnarray}
F_{\theta\varphi} & = & i \sin^2 \gamma \Big[ \frac{1}{2} (\hat{1}+\sigma_z) 
\sin \theta \cos \theta 
\nonumber \\
& & +\sigma _x \sin^2 \theta \cos \varphi \cos \gamma 
\nonumber \\
& &- \sigma_y \sin^2 \theta \sin \varphi \cos \gamma \Big].
\label{eq:exactgaugefield}
\end{eqnarray}
Thus, the gauge field is almost vanishing since $\sin \gamma$ is small for large detuning. 
The associated non-Abelian GP is therefore close to the identity in this limit. 

Second, we prove that the non-Abelian GP cannot be detected since it does not separate 
from the dynamical phase. To do this, we first note that 
the gauge potential in Eq. (\ref{eq:Aexact}) is exact and can therefore be used to calculate formally 
the time evolution operator $U(\tau,0)$ acting on the subspace $\mathcal{S}'$ spanned by the 
exact eigenvectors $\ket{\psi_1}$ and $\ket{\psi_2}$ in Eq. (\ref{eq:exactvectors}), no matter the 
value of $\Delta$. For cyclic evolution of $\mathcal{S}'$, one obtains \cite{anandan88}
\begin{eqnarray}
U(\tau,0) =
{\bf T} e^{-\int_0^{\tau} ( A_{\theta} \dot{\theta} +A_{\varphi} \dot{\varphi}  +
\frac{i}{\hbar} D) dt} ,
\end{eqnarray}
where ${\bf T}$ is time ordering. Here, the dynamical matrix reads
\begin{eqnarray}
D & = &  \frac{1}{2} \hbar \Omega  \tan \gamma \left( \hat{1} + \sigma_z \right) , 
\end{eqnarray}
and $A_{\theta},A_{\varphi}$ are given by Eq. (\ref{eq:Aexact}). One sees that the commutator 
$[D,A_{\theta} \dot{\theta} +A_{\varphi} \dot{\varphi}]$ is on the order of $\sin\gamma$, which 
means that the dynamical and geometric contributions only separate when $\sin \gamma = 0$.  
However, in this case the gauge field in Eq. (\ref{eq:exactgaugefield}) strictly vanishes, and the 
non-Abelian GP and the dynamical phase are both trivial \cite{remark2}. Thus, contrary 
to the claim in Ref. \cite{du11}, a nontrivial non-Abelian GP  ${\bf T} e^{-\int_0^{\tau} ( A_{\theta} 
\dot{\theta} +A_{\varphi} \dot{\varphi} ) dt}$ is not detectable for a path in the space of slow 
parameters $\theta,\varphi$ since it cannot be separated from the dynamical part 
${\bf T} e^{-(i/\hbar) \int_0^{\tau} D dt}$. 

It is instructive to compare the above with the non-Abelian GP proposed in \cite{sjoqvist12} 
for a zero-detuned (i.e., $\tan\gamma = 1$) $\Lambda$ system. In contrast to Ref. \cite{du11}, 
this GP arises in non-adiabatic evolution, as generated by keeping the parameters $\theta$ and 
$\varphi$ fixed, while $\Omega$ is turned on and off so that the subspace spanned by $\ket{1}$ 
and $\ket{2}$ performs a cyclic evolution. The resulting unitary evolution becomes purely geometric 
since $H$ vanishes in this subspace for all times $t$. It can be proved \cite{sjoqvist12} that this 
setting allows for non-commuting GPs that can be used to perform universal quantum computation 
by purely geometric means. 
\vskip 0.3 cm 
M.E. acknowledges support from the Swedish Research Council (VR). E.S. acknowledges 
support from the National Research Foundation and the Ministry of Education (Singapore).  


\begin{thebibliography}{99}
\bibitem{du11} Y.-X. Du, Z.-Y. Xue, X.-D. Zhang, and H. Yan,
Phys. Rev. A {\bf 84}, 034103 (2011).
\bibitem{wilczek84} F. Wilczek and A. Zee,
Phys. Rev. Lett. {\bf 52}, 2111 (1984).
\bibitem{unanyan99} R. G. Unanyan, B. W. Shore, and K. Bergmann,
Phys. Rev. A {\bf 59}, 2910 (1999).
\bibitem{remark} To see this, note that the projection operator $P$ onto $\mathcal{S}$ 
reads $P=\ket{\psi_1}\bra{\psi_1}+\ket{\psi_2}\bra{\psi_2}=\ket{1}\bra{1}+\ket{2}\bra{2}$, 
which is independent of $\theta,\varphi$. 
\bibitem{kult06} D. Kult, J. {\AA}berg, and E. Sj\"oqvist, 
Phys. Rev. A {\bf 74}, 022106 (2006). 
\bibitem{anandan88} J. Anandan, 
Phys. Lett. A {\bf 133}, 171 (1988).
\bibitem{sjoqvist12} E. Sj\"oqvist, D. M. Tong, L. M. Andersson, B. Hessmo,
M. Johansson, and K. Singh,
New J. Phys. {\bf 14}, 103035 (2012).
\bibitem{remark2} One may note that $\sin\gamma=0$ corresponds the Hamiltonian $H = 
2\Delta \ket{3}\bra{3}$, which acts trivially on $\mathcal{S}$. 
\end{thebibliography}
\end{document}